\documentclass[10pt]{iopart}
\usepackage{iopams}
\usepackage{graphicx}
\usepackage{bm}
\begin{document}

\title{Anomalous diffusion and the Moses effect in a model of aging}

\author{Philipp G. Meyer$^1$, Vidushi Adlakha$^{2,3}$, Holger Kantz$^1$ and Kevin E. Bassler$^{2,3,4}$}
\address{$^1$ Max-Planck-Institut f\"{u}r Physik komplexer Systeme, N\"{o}thnitzer Str. 38, Dresden D-01187, Germany}
\address{$^2$ Department of Physics, University of Houston, Houston, TX 77204, USA} 
\address{$^3$ Texas Center for Superconductivity, University of Houston, Houston, TX 77204, USA} 
\address{$^4$ Department of Mathematics, University of Houston, Houston, TX 77204, USA} 
\ead{philippm@pks.mpg.de}

\date{\today}

\begin{abstract}
We decompose the anomalous diffusive behavior found in a model of aging into its fundamental constitutive causes.
The model process is a sum of increments that are iterates of a chaotic dynamical system, the
Pomeau-Manneville map. 
The increments can have long-time correlations, fat-tailed distributions and be non-stationary.
Each of these properties can cause anomalous diffusion through what is known as the Joseph, Noah and Moses effects, respectively. 
The model can have either sub- or super-diffusive behavior, which we find is generally due to a combination of the three effects.
Scaling exponents quantifying each of the three constitutive effects are calculated using analytic methods and confirmed with numerical simulations. They are then related to the scaling of the
distribution of the process through a scaling relation.  
Finally, the importance of the Moses effect in the anomalous diffusion of experimental systems is discussed.
\end{abstract}

\maketitle
\ioptwocol

\section{Introduction}
According to the Central Limit Theorem, the distribution of a process that is the sum of many random increments will have a variance that grows linearly in time. Such processes are said to diffuse normally.
Many experimental systems, however, are known to diffuse anomalously.
Examples include cold atoms in dissipative optical lattices~\cite{Castin1991,Kessler2010}, motion in a crowded environment such as the cytoplasm of biological cells~\cite{Weiss2004,Weiss2013,Ghosh2016}, blinking quantum dots~\cite{Jung2002,MB2004,Sadegh2014}, and intra-day trades in financial markets~\cite{BasslerPNAS07,Seemann2012,ChenPRE17}.
Understanding the nature of the dynamics of these systems that leads to anomalous diffusion is a topic of intense interest.

For stochastic processes that have stationary increments, that is, increments with a time-independent distribution, Mandelbrot~\cite{MandelbrotWaterResouRes68} decomposed the nature of anomalous diffusion into two root causes, or, effects.
He recognized that it can be caused either by long-time increment correlations or by increment distributions that have sufficiently fat tails, so that their variance is infinite.
He called the effect due to increment correlations the \emph{Joseph effect}, and the effect due to fat-tailed increment distributions the \emph{Noah effect}. 
Both effects violate the premises of the Central Limit Theorem.

A third way that the premises of the Central Limit Theorem can be violated is if a stochastic process has non-stationary increments.  Chen, \emph{et al.}, in keeping with the biblically themed names of the other effects, recently named this root cause of anomalous diffusion the \emph{Moses effect}~\cite{ChenPRE17}. 
With the Moses effect, 
the nature of anomalous diffusion in 
processes with non-stationary increments
can now be decomposed as Mandelbrot did for processes with stationary increments. 
Any one of the Joseph, Noah or Moses effects, or a combination of
them, can cause anomalous diffusion.

For self-affine processes, which have a distribution that scales with a power-law of time $t^H$, where $H$ is what Mandelbrot called the \emph{Hurst exponent},
scaling exponents can be defined to quantify each of the three effects that can cause anomalous scaling.
The \emph{Joseph exponent} $J$ quantifies the increment correlations in the Joseph effect.
When $J \neq 1/2$ increment correlations exist and can cause anomalous scaling.
If $J > 1/2$ the increments are positively correlated, while if $J < 1/2$ they are anti-correlated.
The \emph{latent exponent} $L$ quantifies the effect of increment distribution fat-tails in the Noah effect. 
When $L > 1/2$, the increment distribution has ``fat tails" and anomalous scaling can result.
The \emph{Moses exponent} $M$ quantifies the effect of non-stationarity of the increment distributions in the Moses effect.
When $M > 1/2$ the increment distribution widens with time, and for $M<1/2$ it shrinks with time.
The exponents $J$, $L$ and $M$ are related to $H$ through the scaling relation
\begin{equation}
H = J+L+M-1 \; .
\label{HJLM}
\end{equation}
If $H \neq 1/2$ the diffusion is anomalous. If $H < 1/2$ the process is sub-diffusive, and if $H > 1/2$ it is super-diffusive.
Robust statistical methods that analyze ensembles of realizations of a stochastic process can be used to independently determine each of the four exponents~\cite{ChenPRE17}. 

Physical systems with aging behavior can have non-stationary, time-dependent behavior and diffuse anomalously. 
Aging systems can also be non-Markovian, having long-time increment correlations, which 
also can contribute the anomalous behavior~\cite{BarkaiJCP2003,Schulz2014,Safdari2015}.
In this paper, we decompose the anomalous diffusion found in a simple model of aging behavior~\cite{Geisel1984, BarkaiPRL03} and find that it
is due to a rich combination of the Joseph, Noah and Moses effects. The model process consists of increments that 
are generated by a nonlinear map. Although the map is a deterministic system, it is intermittently chaotic, and the increments it 
generates model stochastic, noise driven increments in physical systems. 

The paper is organized as follows.
In the next section we introduce our model system and describe its anomalous diffusive behavior.
Then, in the third section, we quantitatively decompose the anomalous diffusion into its root effects using analytic scaling arguments for each of the different constitutive effects. 
In the forth section, we confirm our analytic results with numerical simulations.
In the final section, we discuss our results and the importance of the Moses effect for anomalous diffusive behavior observed in experimental systems.

\section{Model and its Diffusive Behavior}
Consider a one-dimensional, discrete-time process $X_t$, defined for integer time $t \in [0,\infty)$.
The process is the sum of increments $\{\delta_t\}$
\begin{equation}
X_t = \sum_{s=0}^{t-1} \delta_s \; ; \; X_0 = 0
\end{equation}
that are iterates of the modified Pomeau-Manneville (PM) map~\cite{PomeauCommMathPhys80}
\begin{equation}
\delta_{t+1}
=
\left\{
\begin{array}{ll}
-4\delta_t + 3 & \mbox{if} \; 0.5 < \delta_t \leq 1.0 \\
\delta_t\left(1+\left|2\delta_t\right|^{z-1}\right) & \mbox{if} \; |\delta_t| \leq 0.5 \\
-4\delta_t - 3 & \mbox{if} \; -1 \leq \delta_t < -0.5
\end{array}
\right. \; .
\end{equation}
This map has been studied extensively in the past. It has been linked to anomalous diffusion \cite{Geisel1984} aging \cite{BarkaiPRL03} and weak ergodicity breaking \cite{Bel2006}.
The initial increment $\delta_0$ is chosen randomly from a uniform distribution in the interval $\left[ -1,1 \right]$. 

\begin{figure}[ptb]
\centering
\includegraphics[width=0.44\textwidth]{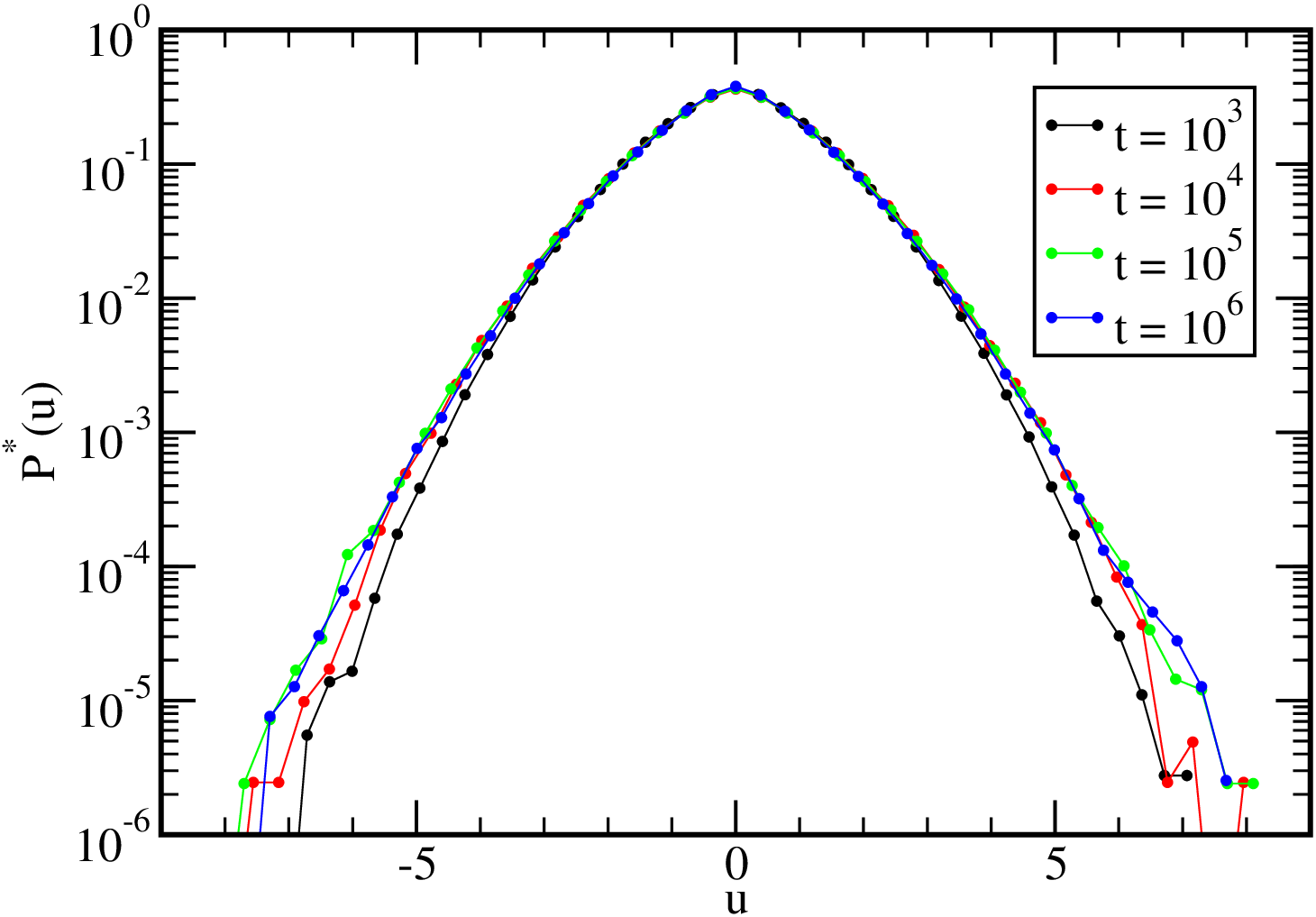}
\caption{Scaling function of the distribution of the process $X$. The scaling parameter is $u = X_t/t^H$. The function is shown at four different times $t$, spanning 3 decades from $10^3$ to $10^6$. Here the PM map parameter is $z=2.5$ and the Hurst exponent is $H = 0.38$. Lines connecting the data points are shown as guides to the eye.}
\label{fig1}
\end{figure}

The distribution of the process $P(X_t)$ scales with time
\begin{equation}
P(X_{t}) = t^H P^*\!\left(X_t/t^H\right) 
\end{equation}
where $H$ is the \emph{Hurst exponent} and $P^*$ is the \emph{scaling function} shown in Fig.~\ref{fig1}. 
In the figure, the scaling functions converge for large $t$, but corrections to scaling are noticeable at smaller $t$.
To measure $H$ empirically, one can simply measure the scaling of the width of distribution
\begin{equation}
w[X_t] \sim t^H \; , \label{Xscaling}
\end{equation}
which can be defined as, say, the distance between the 75th quantile and the 25th quantile of $P(X_t)$, or, 
as the standard deviation of the distribution $\sqrt{\langle X_t^2 \rangle}$, if it is finite.  
Here $\langle X_t \rangle = 0$ for all $t$.

The value of $H$ can be calculated from the theory of stochastic renewal processes. A more rigorous calculation was presented in \cite{NiemannDis}. Here we recall the salient points. Stochastic renewal processes have a well defined waiting time distribution. In our case a waiting period is defined as the time that the system spends in the interval $|\delta| \leq 0.5$. In this region the dynamics of the system is regular with $\delta$ monotonically increasing (or decreasing if it's negative), while it escapes from the unstable fixed point at $\delta=0$.
The dynamics can be approximated by a continuous differential equation
\begin{equation}
\label{delta->0}
{{\rm d} \delta \over {\rm d} t} = \delta \; (2 \delta)^{z-1} \  \ \mbox{when} 
\ \delta \rightarrow 0.
\end{equation}
Once $\delta$ reaches the outer part $|\delta| > 0.5$, the dynamics is chaotic until a ``reinjection" into the regular region occurs. Hence, this type of system is intermittently chaotic. Generally the chaotic motion is very short, thus
it is sufficient to consider only the waiting periods.

Integrating Eq.~(\ref{delta->0}) and assuming a uniform distribution for reinjected $\delta$, the distribution of waiting times $\tau$ can be calculated \cite{Geisel1984,Akimoto2014}
\begin{equation}
\label{wtd}
P({\tau}) \propto {\tau}^\frac{z}{1-z}.
\end{equation}
For $z>2$ the mean waiting time diverges, which gives rise to all the effects we discuss in this article.
During each waiting period the process $X_t$ performs a ``jump" due to many small steps $\delta$ in the same direction. We define the displacement $\chi$ during jumps that start at renewal (reinjection) times $t=t_r$ as
\begin{equation}
\chi = X_{t_r+\tau}-X_{t_r}\; .
\end{equation}
It can be calculated from Eqs. (\ref{delta->0},\ref{wtd}) to be~\cite{MeyerPRE17} 
\begin{equation}
\chi \sim \frac{1}{2}(z-1)^\frac{2-z}{1-z}\frac{1}{z-2}\tau^\frac{2-z}{1-z},
\end{equation}
thus giving a relation between $\chi$ and the waiting times. The joint probability density function (PDF) for $\chi$ and $\tau$ then is
\begin{equation}
\label{jointPDF}
P(\chi,\tau) \propto {\tau}^\frac{z}{1-z} {1 \over 2} \left[ \delta(\chi-\tau^\frac{2-z}{1-z}) 
+ \delta(\chi + \tau^\frac{2-z}{1-z}) \right],
\end{equation}
where $\delta$ here is the Dirac $\delta$-function.
Eq.~(\ref{jointPDF}) is well known in stochastic renewal processes \cite{Metzler2000}.

From here the value of the Hurst exponent can be calculated from the scaling of the ensemble averaged mean squared displacement of the process $\langle X_t^2\rangle\propto t^{2H}$.
As for continuous time random walks~\cite{AkimotoMiyaguchi2014}, the calculation can be performed in Fourier-Laplace space. There the mean squared displacement is equal to the second derivative of the spatial distribution of the process.
Results are known for processes of with PDFs of the form of Eq.~(\ref{jointPDF}) for Levy flights \cite{AkimotoMiyaguchi2014} and Levy walks \cite{AlbersPRL}. The results in both cases are the same and therefore independent of the exact path the system takes during each waiting time. They thus apply to our system as well~\cite{MeyerPRE17,Meyer2017b,NiemannDis}
\begin{equation}
\label{HofT}
H =
\left\lbrace\begin{array}{l l}
0.5 & \mbox{if }z<2\\
0.5/(z-1) & \mbox{if }2<z<\frac{5}{2}\\
(z-2)(z-1) & \mbox{if }\frac{5}{2}<z
\end{array}\right. .
\end{equation}
At the crossover points logarithmic corrections appear \cite{Akimoto2014}, which make numerical calculations more difficult.
For $2<z<3$, $H < \frac12$ and the system is sub-diffusive, while for $z>3$, $H > \frac12$ and the system is super-diffusive.

\section{Joseph, Noah and Moses effects}
\subsection{Definitions}
Anomalous diffusion can be decomposed into effects that are root causes for the violation of the premises of the CLT. In order for the CLT to hold for a process $X_t$ that is the sum of random increments $\{\delta_t\}$, the increments must: (1) be independent, (2) have a distribution with a finite variance, and (3) be identically distributed. Violation of these premises is referred as the Joseph, Noah, and Moses effects, respectively.  Each of these constitutive effects, or a combination of them, can cause anomalous diffusion. For self-affine processes, they can be quantified by scaling exponents, which are related to each other and the Hurst exponent by Eq.~\ref{HJLM}. These exponents are defined as follows.

To define the exponents, first define the following random variables:
the sum of the absolute values of increments
\begin{equation}
Y_t = \sum\limits_{s=0}^{t-1}|\delta_s| ,
\label{Y}
\end{equation}
and the sum of increment squares
\begin{equation}
Z_t = \sum\limits_{s=0}^{t-1} \delta^{2}_s.
\label{Z}
\end{equation}
Then, the Moses exponent $M$ and the Latent exponent $L$, which quantify the Moses effect and Noah effect, respectively, are defined by the scaling of ensemble-averaged median of these variables
\begin{eqnarray}
m[Y_t] & \sim & t^{M+1/2} \label{Yscaling} \\
m[Z_t] & \sim & t^{2L+2M-1} \; , \label{Zscaling}
\end{eqnarray}
or, similarly, of their means, if they are finite.

The Joseph exponent $J$, which quantifies the Joseph effect, is defined by the scaling of the ensemble-averaged rescaled range statistic (R/S)~\cite{Hurst51}  
\begin{equation}
E[R_t/S_t] \sim t^J \; , \label{RSscaling}
\end{equation}
where $R_t$ is the range of the process
\begin{equation}
R_t  =\max\limits_{1 \leq s \leq t}\left[X_s - \frac{s}{t}X_t\right] 
- \min\limits_{1 \leq s \leq t}\left[X_s - \frac{s}{t}X_t\right]
\end{equation}
and $S_t$ is the standard deviation of increments up to time $t$
\begin{equation}
S^{2}_t = \frac{1}{t} Z_t - \left[\frac{1}{t}X_t\right]^2 \; .
\end{equation}

It should be noted that there is some confusion in the literature with exponents $J$ and $H$.
The $J$ defined in Eq.~\ref{RSscaling} is the exponent originally defined by Hurst,
and is the exponent that is found through detrended fluctuation analysis (DFA)~\cite{Peng1994,Hoell2015}. 
In papers utilizing DFA it is often referred to as the Hurst exponent, e.g. in \cite{Kantelhardt2002}. 
Mandelbrot first called this exponent $J$ and distinguished it from the $H$ 
defined in Eq.~\ref{Xscaling}~\cite{MandelbrotWaterResouRes68}.
Of course, in processes with no Noah or Moses effects, $J$ and $H$ become equivalent.

\subsection{Values in the Pomeau-Manneville map}

In order to calculate the values of the scaling exponents for the PM map, we utilize the concept of infinite invariant densities in infinite ergodic theory. Such densities have well defined shape that scale with time.
The increment density $P(\delta_t)$ is a good example of this. It does not satisfy the central limit theorem. However, it does satisfy two other limit theorems.

First, it can be shown \cite{Korabel2009}, that there exists an infinite invariant density $P_{\rm \inf}(|\delta|)\propto |\delta|^{1-z}$ 
that is related to the actual density by
\begin{equation}
\label{inf}
P(|\delta_t|) \sim t^\frac{2-z}{z-1} P_{\rm \inf}(|\delta|) 
\end{equation}
for $|\delta_t|$ not close to zero.
The density $P(|\delta_t|)$ must be normalizable since it is a physical density.
Therefore, it is truncated for small values of $\delta$. For longer aging times $t$, the location of the cutoff of the density moves closer and closer towards zero. Thus, the infinite invariant density in the limit of large $t$ has a well defined power law shape, but is not integrable.

There is a second way to obtain an invariant expression related to the increment density. It was derived by Dynkin \cite{Dynkin} in the context of renewal theory, while Thaler \cite{Thaler2005} established the
connection to the underlying transformations. 
The application to the PM map was shown in \cite{AkimotoBarkai}.
It can be shown that by transforming the increment $\delta$ according to
\begin{equation}
\label{gamma}
\gamma=2|\delta|\left(t(z-1)\right)^\frac{1}{z-1} ,
\end{equation}
one obtains an invariant distribution
\begin{equation}
\label{Pgamma}
P(\gamma) =\frac{z-1}\pi\sin\left(\frac{\pi}{z-1}\right)\frac1{1 + \gamma^{z-1}}
\end{equation}
for $\gamma$.

Both of these limit theorems are necessary for understanding the scaling of $Z_t$.
It is calculated by
\begin{equation}
\langle Z_t\rangle=\sum_{s=1}^t \langle \delta_{s}^2\rangle=\sum_{s=1}^t\int_0^1\mathrm{d}|\delta_{s}| \ \delta_{s}^2 P(|\delta_{s}|).
\end{equation}
The sum just adds a '+1' to the exponent describing the scaling behavior of $\langle \delta_s^2 \rangle$, which was calculated in \cite{MeyerPRE17}. For $z<2$ the scaling is trivial, because $P(\delta)$ is stationary and integrable. For $2<z<4$, it can be found using the infinite invariant density in Eq. (\ref{inf}).  For $z>4$ the integral over $\delta^2 P_{\mathrm{inv}}(|\delta|)$ diverges, but the Thaler-Dynkin limit theorem can be applied using Eq. (\ref{gamma}).
From these considerations $Z_t$ scales as:
\begin{equation}
\langle Z_t\rangle \propto
\left\lbrace\begin{array}{l l}
t & \mbox{if }z<2\\
t^\frac{1}{z-1} & \mbox{if }2<z<4\\
t^\frac{z-3}{z-1} & \mbox{if }4<z
\end{array}\right. .
\end{equation}

\begin{figure}[ptb]
\centering
\includegraphics[width=0.44\textwidth]{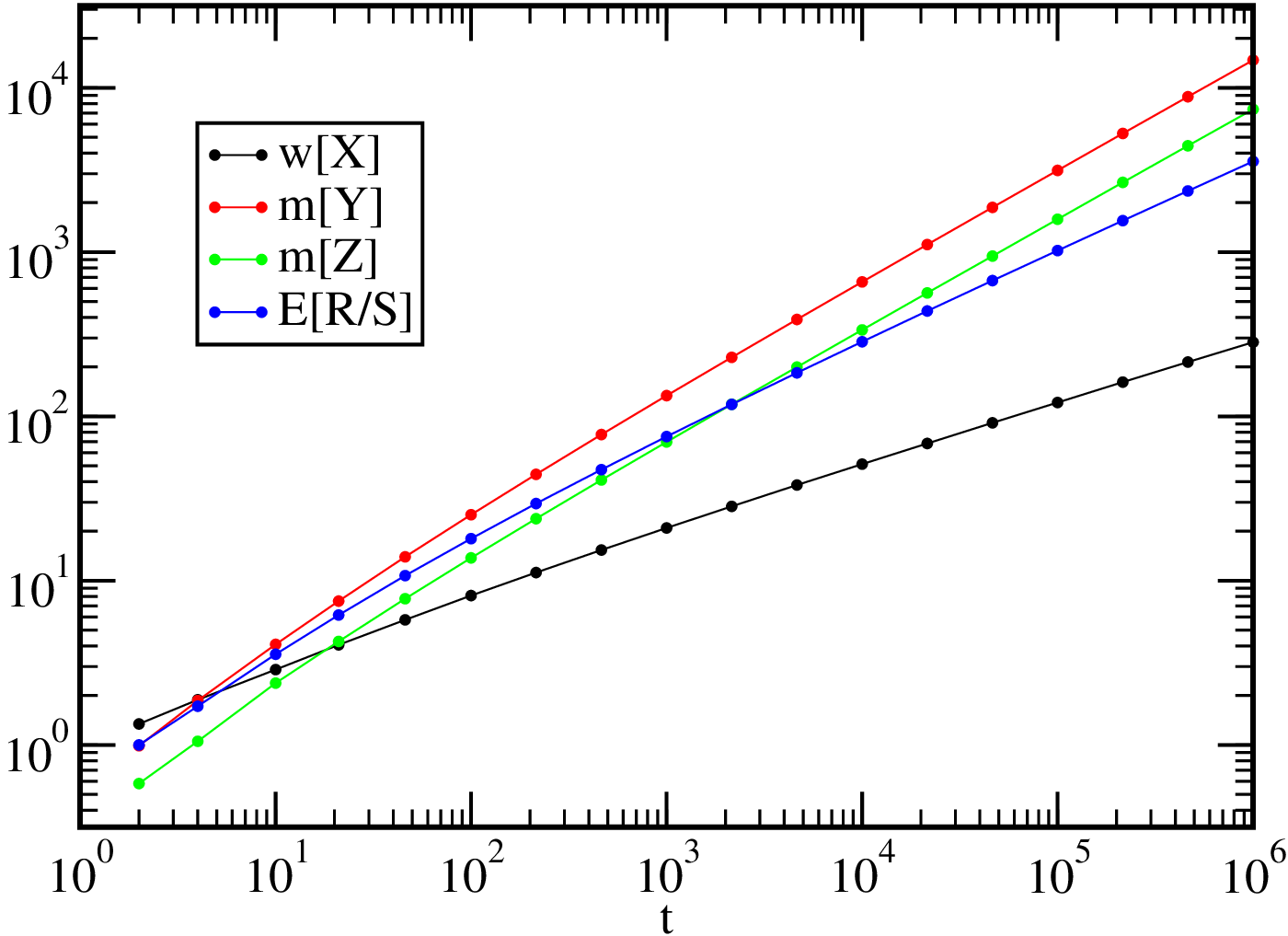}
\caption{Log-log plot of the width of $X$, median of $Y$ and of $Z$, and mean of $R/S$ as a function of time $t$. The scaling of these functions, quantified by the exponents corresponding to the slope of the function at large $t$, determine the exponents quantifying diffusion and the constitutive effects that cause anomalous diffusion. Here the PM map parameter is $z = 2.5$. Lines connecting the data points are shown as guides to the eye.}
\label{fig2}
\end{figure}

Similar methods can be used to find the scaling behavior of $Y_t$. Again, we use the ensemble average
\begin{equation}
\label{ZofT}
\langle Y_t\rangle=\sum_{s=1}^t \langle |\delta_{s}|\rangle=\sum_{s=1}^t\int_0^1\mathrm{d}|\delta_{s}| \ |\delta_{s}| P(|\delta_{s}|).
\end{equation}
Here  the expression is integrable with respect to the infinite invariant density for $z<3$ and with respect to the Thaler-Dynkin limit theorem for $z>3$. Therefore, $\langle Y_t \rangle$ scales as
\begin{equation}
\label{YofT}
\langle Y_t\rangle \propto
\left\lbrace\begin{array}{l l}
t & \mbox{if }z<2\\
t^\frac{1}{z-1} & \mbox{if }2<z<3\\
t^\frac{z-2}{z-1} & \mbox{if }3<z
\end{array}\right. .
\end{equation}
Note that the scaling of the mean is equivalent to the scaling of the median for both $Y_t$ and $Z_t$.

No more information is needed to calculate the exponents $J$, $L$ and $M$. Using Eqs.(\ref{Yscaling}), (\ref{Zscaling}), (\ref{ZofT}) and (\ref{YofT}), one obtains for the Moses exponent
\begin{equation}
\label{MofT}
M =
\left\lbrace\begin{array}{l l}
0.5 & \mbox{if }z<2\\
(1.5-0.5z)/(z-1) & \mbox{if }2<z<3\\
(0.5z-1.5)/(z-1) & \mbox{if }3<z
\end{array}\right. ,
\end{equation}
and for the latent exponent
\begin{equation}
\label{LofT}
L =
\left\lbrace\begin{array}{l l}
0.5 & \mbox{if }z<2\\
(z-1.5)/(z-1) & \mbox{if }2<z<3\\
1.5/(z-1) & \mbox{if }3<z<4\\
0.5 & \mbox{if }4<z
\end{array}\right. .
\end{equation}
Since the Hurst exponent $H$ is given by Eq. (\ref{HofT}), the Joseph exponent $J$ can be determined using the scaling relation Eq.~(\ref{HJLM}),
\begin{equation}
\label{JofT}
J =
\left\lbrace\begin{array}{l l}
0.5 & \mbox{if }z<2.5\\
(1.5z-3)/(z-1) & \mbox{if }2.5<z<4\\
1 & \mbox{if }4<z
\end{array}\right. .
\end{equation}

At the end of this section we want to add a short remark about the parameter $J$. Throughout this section followed a similar path of reasoning as in \cite{MeyerPRE17}, where the ensemble averaged time averaged mean squared displacement (EATAMSD) of the PM map was shown to be
\begin{equation}
\left\langle \overline{X^2}\right\rangle \equiv\sum_{s=0}^{t-\Delta}\frac{(X_{s+\Delta}-X_{s})^2}{t-\Delta}\propto t^\beta \Delta^{2H-\beta},
\end{equation}
for $z>2.5$. Here $t$ is the total measurement time.
This is the result of the scale-invariant Green-Kubo relation for time averaged diffusivity. 
The scaling of the EATAMSD is the difference between the scaling of the ensemble averaged mean squared displacement ($=2H$) and the scaling of the "velocity displacement" $\langle \delta_t^2\rangle\propto t^\beta$, which is in fact the derivative of $Z_t$. So $Z_t$ scales like $t^{\beta+1}$.
Then considering Eq. (\ref{Zscaling}), the scaling exponent of the EATAMSD in fact is $2H-2L-2M+2$. This result from the scale-invariant Green-Kubo relation for time averaged diffusivity looks very much like our scaling relation Eq.~(\ref{HJLM}) and therefore implies that the EATAMSD scales as $\sim t^{2J}$. This equivalence is true, at least, for systems with scale invariant increment correlation functions $\langle\delta_{t}\delta_{t+\Delta}\rangle\propto t^{2H-2}\Phi (\Delta /t)$ in the parameter range $J>0.5$ and $L+M>0.5$.

\begin{figure}[ptb]
\centering
\includegraphics[width=0.40\textwidth]{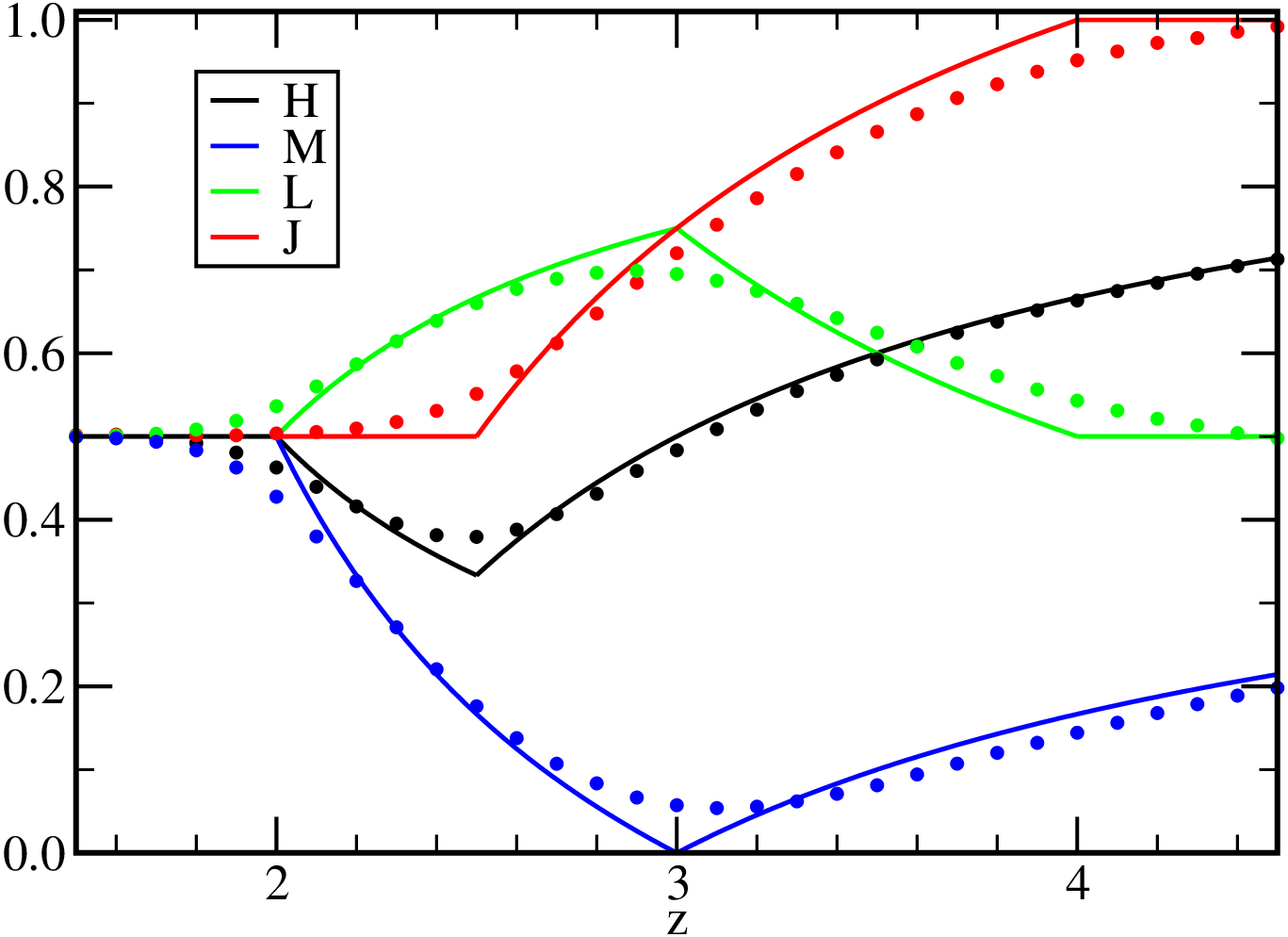}
\caption{Values of the scaling exponents quantifying diffusion and the constitutive effects that cause anomalous diffusion in the PM map as a function of parameter $z$. Circles show the values resulting from numerical simulation and the solid lines the analytic predictions. As $z\rightarrow \infty$, $H \rightarrow 1$ and $M \rightarrow \frac12$. 
}
\label{fig3}
\end{figure}

\section{Simulation results}
To verify our analytic predictions for the exponents, we performed numerical simulations of the PM map.
For each value of $z$ from 1.5 to 4.5 in steps of 0.1, we generated an ensemble of $10^5$ realizations of the process $X_t$ for $t=10^6$ map iterations. We then measured $w[X_t]$, $m[Y_t]$, $m[Z_t]$, and $E[R_t/S_t]$ for the ensemble. Example results for $z=2.5$ are shown in Fig.~\ref{fig2}. The statistical error of the data points is smaller than the symbol size. The scaling exponents describe the asymptotic, large $t$ scaling behavior of these functions. 

\begin{figure}[ptb]
\centering
\includegraphics[width=0.40\textwidth]{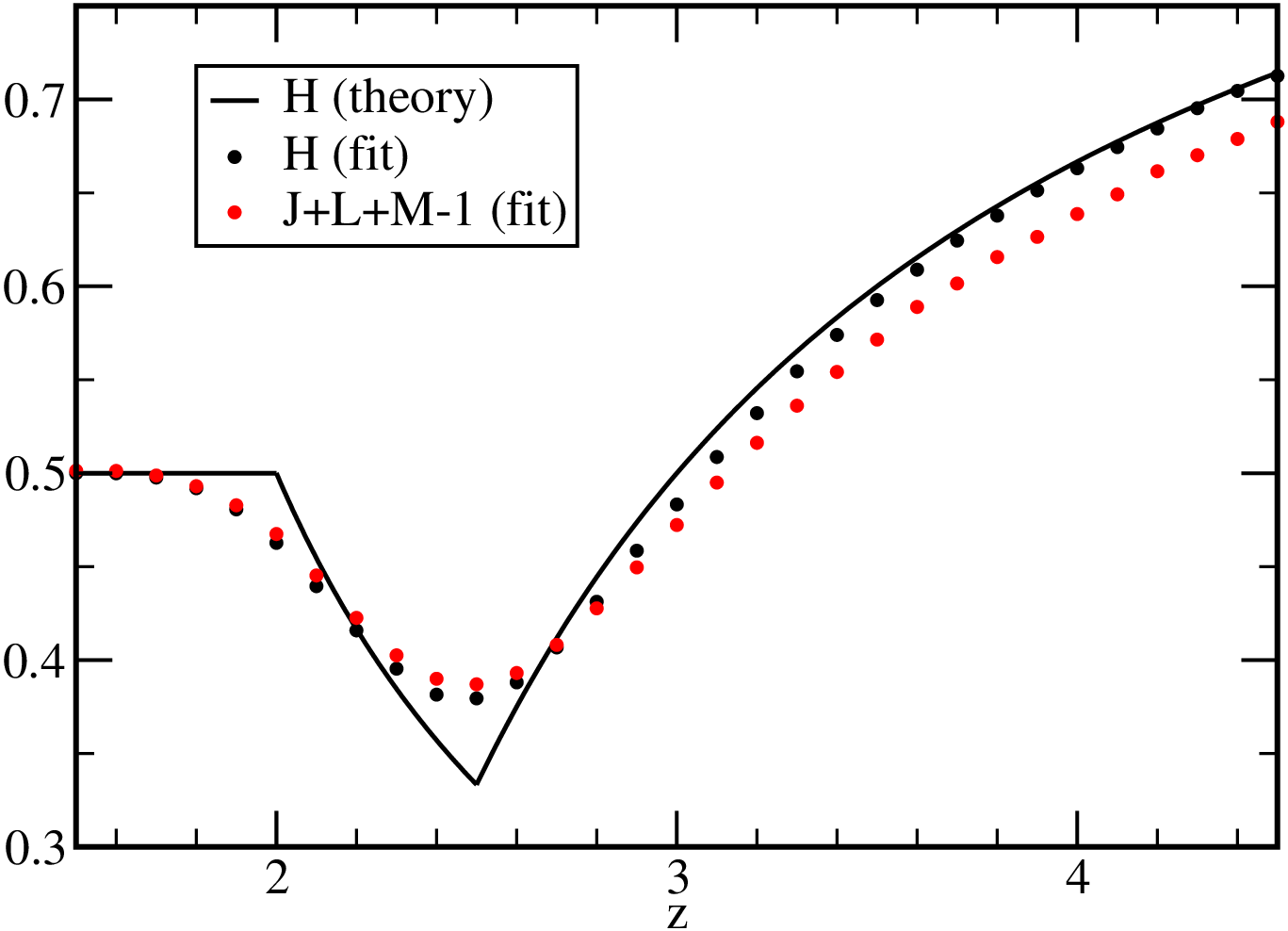}
\caption{Verification of the scaling relation (Eq.~\ref{HJLM}) for the exponents quantifying diffusion in the PM map as a function of parameter $z$.
}
\label{fig4}
\end{figure}

We fit each of the four functions over the two-decade range from $t=10^4$ to $10^6$ to the form
\begin{equation}
f(t) = at^{\Omega}
\end{equation}
where $a$ and $\Omega$ are fitting parameters, 
and $\Omega$ is the asymptotic scaling exponent.
Finding $\Omega$ for each function, we then determined the values of $H$, $M$, $L$ and $J$ using Eqs.~\ref{Xscaling}, \ref{Yscaling}, \ref{Zscaling} and \ref{RSscaling}. The result of the fitting for each for each exponent as a function of $z$ is shown by the filled circles in Fig.~\ref{fig3}. The analytic predictions are shown by the solid lines in the figure.
Fig.~\ref{fig4} confirms the scaling relation between the exponents, Eq.~\ref{HJLM}. It shows the theoretically predicted 
value of $H$, versus the fitted value of $H$ from simulations and the result of the fitting for $J+L+M-1$ as a function of $z$. 

The simulation results roughly follow our theoretical predictions, but there are deviations. 
Upon close examination of Fig.~\ref{fig2}, one can see that the functions are still curving on a log-log plot at $t=10^6$. 
Thus the deviations between simulation results and their predictions in Figs.~\ref{fig3} and \ref{fig4} are
presumably due to finite-time corrections to scaling.
The corrections are especially evident near crossover points where they have a
logarithmic form~\cite{Akimoto2014}.
To obtain more accurate numerical results, either corrections to scaling must be included in the fit 
or longer simulations must be run.
The general form of the corrections, however, are not known, and
the accuracy of long simulations are limited by the phenomenon of ``round-off periodicity"~\cite{Beck1987,Grebogi1988}. With double precision calculations, the map can accurately be iterated only about $10^6$ times. Longer simulations would require computationally expensive quadruple precision calculations.

\section{Discussion}
The root causes of anomalous diffusion can be decomposed into the Joseph, Noah and Moses effects. 
Nonstationary increments, such as what occurs in aging processes, can cause anomalous diffusion through the Moses effect. Previous studies have found that 
the anomalous diffusion in intra-day financial market prices is solely due to the Moses effect~\cite{BasslerPhysicaA06,BasslerPNAS07,McCauleyPhysicaA07,McCauleyPhysicaA08,Seemann2012,ChenPRE17}. 
Here, however, we have found that the anomalous diffusion in a simple model of aging is due to a complex combination of the Joseph, Noah and Moses effects.

It is perhaps surprising that a process consisting of increments that are iterates of the PM map can have such a rich set of behaviors. The PM map is, after all, a deterministic map from the interval $[-1,1]$ onto itself. The value of increments, thus, is bounded. The dynamics of the map, however, consists of an intermittent mixture of regular and chaotic motion,
which can produce anomalous diffusion. 
For $z \leq 2$, nothing surprising happens. The diffusion of the system is normal, the increments are stationary, and there is no Joseph, Noah or Moses effect.
But for $z>2$, the system ages with time, because the mean waiting time in the  regular region diverges. This causes the increment distribution to become time-dependent and a sub-diffusive Moses effect to occur.

Remarkably, a Noah effect occurs despite the fact that the tails of the increment distribution are bounded.
However, for $z>2$, where the increment distribution is time-dependent, rescaling of $\delta$ using Eq.~(\ref{gamma}) 
results in a stationary increment distribution that has fat tails!
The second moment of this stationary distribution Eq.~(\ref{Pgamma}) diverges for $2<z<4$, causing the Noah effect. For larger values of $z$, although the distribution still has power law tails, they decay fast enough to keep its variance finite. 

For $z<2$ the density $P(|\delta|)$ is stationary and has a well defined mean. For $2<z<3$ the density becomes nonstationary and moves towards zero. However, since it has fat tails, the mean goes to zero more slowly than the density itself. For $z>3$, the distribution (Eq. (\ref{Pgamma})) is steeper. Here the mean goes to zero with the same scaling as the distribution itself, causing a decrease of the Moses effect. As a consequence the Noah effect also decreases, because it is directly linked to the Moses effect, as stated before.

For $z<2.5$, there are no long-time increment correlations, as they are disrupted by the frequent intermittent periods of chaotic motion. However, for $z> 2.5$, increment correlations do exist and contribute to the anomalous diffusive behavior through the Joseph effect, which increases with $z$ and becomes maximal at $z=4$. 
Both the Noah and the Joseph effects only lead to super-diffusive behavior in the system, as both $L$ and $J \geq 0.5$ for all $z$, while the Moses effect only leads to sub-diffusive behavior, as $M \leq 0.5$ for all $z$. The three effects combine to produce sub-diffusive behavior in the model for $2 < z < 3$. In this range of $z$, the Moses effect dominates over the other two effects. For $z > 3$, super-diffusive behavior occurs instead as the Noah and Joseph dominate the Moses effect.    

In empirical time-series analyses it is often assumed that the increments of the process are stationary~\cite{Mandelbrot1963,MantegnaStanley1995}.
This assumption is made to justify "sliding-window" statistical analyses that combine different increments.
This can lead, however, to spurious results if the increments are non-stationary, 
such as falsely determining that the process has a fat-tailed distribution~\cite{BasslerPNAS07}.
If there are non-stationary increments, a proper statistical analysis requires studying an
ensemble of processes. This can be difficult to acquire such data, especially if the data is a historical time-series. 
For some systems, there may be periodic or intermittent triggering events that can be thought of as beginning a new process~\cite{BasslerPNAS07,ChenPRE17}.
For others, experiments must be repeated. Here we simply repeated the numerical simulations to acquire the
data for the ensemble analysis.

We have shown that decomposing anomalous diffusive behavior into its fundamental constitutive causes can 
contribute to understanding the nature of the system's dynamics.
It would be interesting to similarly decompose the anomalous diffusive behavior found in other systems, especially experimental systems. 
For other aging systems, which by definition have non-stationary increments, such as blinking quantum dots~\cite{Jung2002,MB2004,Sadegh2014},  it can be expected that the Moses effect contributes to anomalous diffusive behavior, but do the Joseph and Noah effects also contribute to the observed behavior?
In other systems that diffuse anomalously, which are not necessarily known as aging systems, will a
decomposition of the diffusion into constitutive effects reveal that a Moses effect contributes to the observed behavior?

\ack
This research is supported in part by the US National Science Foundation, through grant DMR-1507371 (VA and KEB).
KEB also thanks the Max-Planck-Institut f\"{u}r Physik komplexer Systeme for its support and hospitality during his visit when this work was initiated.

\section*{References}
\bibliography{aging}

\begin{thebibliography}{10}

\bibitem{Castin1991}
Y.~Castin, J.~Dalibard, and C.~Cohen-Tannoudji.
\newblock The limits of sisyphus cooling.
\newblock In L.~Mol, S.~Gozzini, C.~Gabbanini, E.~Arimondo, and F.~Strumia,
  editors, {\em Light Induced Kinetic Effects on Atoms, Ions, and Molecules},
  pages 5--24, Pisa, 1991. ETS Editrice.

\bibitem{Kessler2010}
David~A. Kessler and Eli Barkai.
\newblock Infinite covariant density for diffusion in logarithmic potentials
  and optical lattices.
\newblock {\em Phys. Rev. Lett.}, 105:120602, Sep 2010.

\bibitem{Weiss2004}
Matthias Weiss, Markus Elsner, Fredrik Kartberg, and Tommy Nilsson.
\newblock Anomalous subdiffusion is a measure for cytoplasmic crowding in
  living cells.
\newblock {\em Biophysical Journal}, 87:3518--3524, 2004.

\bibitem{Weiss2013}
Matthias Weiss.
\newblock Single-particle tracking data reveal anticorrelated fractional
  brownian motion in crowded fluids.
\newblock {\em Phys. Rev. E}, 88:010101, Jul 2013.

\bibitem{Ghosh2016}
Surya~K Ghosh, Andrey~G Cherstvy, Denis~S Grebenkov, and Ralf Metzler.
\newblock Anomalous, non-gaussian tracer diffusion in crowded two-dimensional
  environments.
\newblock {\em New Journal of Physics}, 18(1):013027, 2016.

\bibitem{Jung2002}
YounJoon Jung, Eli Barkai, and Robert~J. Silbey.
\newblock Lineshape theory and photon counting statistics for blinking quantum
  dots: a lévy walk process.
\newblock {\em Chemical Physics}, 284(1):181 -- 194, 2002.
\newblock Strange Kinetics.

\bibitem{MB2004}
G.~Margolin and E.~Barkai.
\newblock Aging correlation functions for blinking nanocrystals, and other
  on–off stochastic processes.
\newblock {\em The Journal of Chemical Physics}, 121(3):1566--1577, 2004.

\bibitem{Sadegh2014}
Sanaz Sadegh, Eli Barkai, and Diego Krapf.
\newblock 1/ f noise for intermittent quantum dots exhibits non-stationarity
  and critical exponents.
\newblock {\em New Journal of Physics}, 16(11):113054, 2014.

\bibitem{BasslerPNAS07}
Kevin~E. Bassler, Joseph~L. McCauley, and Gemunu~H. Gunaratne.
\newblock Nonstationary increments, scaling distributions, and variable
  diffusion processes in financial markets.
\newblock {\em Proceedings of the National Academy of Sciences},
  104(44):17287--17290, 2007.

\bibitem{Seemann2012}
Lars Seemann, Jia-Chen Hua, Joseph~L. McCauley, and Gemunu~H. Gunaratne.
\newblock Ensemble vs. time averages in financial time series analysis.
\newblock {\em Physica A: Statistical Mechanics and its Applications},
  391(23):6024 -- 6032, 2012.

\bibitem{ChenPRE17}
Lijian Chen, Kevin~E. Bassler, Joseph~L. McCauley, and Gemunu~H. Gunaratne.
\newblock Anomalous scaling of stochastic processes and the moses effect.
\newblock {\em Phys. Rev. E}, 95:042141, Apr 2017.

\bibitem{MandelbrotWaterResouRes68}
Mandelbrot~Benoit B. and Wallis~James R.
\newblock Noah, joseph, and operational hydrology.
\newblock {\em Water Resources Research}, 4(5):909--918.

\bibitem{BarkaiJCP2003}
Eli Barkai and Yuan-Chung Cheng.
\newblock Aging continuous time random walks.
\newblock {\em J. Chem. Phys.}, 118:6167, Apr 2003.

\bibitem{Schulz2014}
Johannes H.~P. Schulz, Eli Barkai, and Ralf Metzler.
\newblock Aging renewal theory and application to random walks.
\newblock {\em Phys. Rev. X}, 4:011028, 2014.

\bibitem{Safdari2015}
Hadiseh Safdari, Aleksei~V. Chechkin, Gholamreza~R. Jafari, and Ralf Metzler.
\newblock Aging scaled brownian motion.
\newblock {\em Phys. Rev. E}, 91:042107.

\bibitem{Geisel1984}
Theo Geisel and Stefan Thomae.
\newblock Anomalous diffusion in intermittent chaotic systems.
\newblock {\em Phys. Rev. Lett.}, 52:1936, 1984.

\bibitem{BarkaiPRL03}
Eli Barkai.
\newblock Aging in subdiffusion generated by a deterministic dynamical system.
\newblock {\em Phys. Rev. Lett.}, 90:104101, Mar 2003.

\bibitem{PomeauCommMathPhys80}
Yves Pomeau and Paul Manneville.
\newblock Intermittent transition to turbulence in dissipative dynamical
  systems.
\newblock {\em Communications in Mathematical Physics}, 74(2):189--197, Jun
  1980.

\bibitem{Bel2006}
Golan Bel and Eli Barkai.
\newblock Ergodicity breaking in a deterministic system.
\newblock {\em Europhys. Lett.}, 74:15, 2006.

\bibitem{NiemannDis}
Markus Niemann.
\newblock From anomalous deterministic diffusion to the continuous-time random
  walk.
\newblock {\em Ph.D. diss.}, 2009.

\bibitem{Akimoto2014}
Takuma Akimoto, Soya Shinkai, and Yoji Aizawa.
\newblock Distributional behavior of time averages of non-l1 observables in
  one-dimensional intermittent maps with infinite invariant measures.
\newblock {\em J. Stat. Phys.}, 158:476, 2014.

\bibitem{MeyerPRE17}
Philipp Meyer, Eli Barkai, and Holger Kantz.
\newblock Scale-invariant green-kubo relation for time-averaged diffusivity.
\newblock {\em Phys. Rev. E}, 96:062122, Dec 2017.

\bibitem{Metzler2000}
Ralf Metzler and Joseph Klafter.
\newblock The random walk's guide to anomalous diffusion: a fractional dynamics
  approach.
\newblock {\em Phys. Rep.}, 339:1--77, 2000.

\bibitem{AkimotoMiyaguchi2014}
Takuma Akimoto and Tomoshige Miyaguchi.
\newblock Phase diagram in stored-energy-driven levy flight.
\newblock {\em J. Stat. Phys.}, 157:515, 2014.

\bibitem{AlbersPRL}
Tony Albers and Günter Radons.
\newblock Exact results for the nonergodicity of d-dimensional generalized levy
  walks.
\newblock {\em Phys. Rev. Lett.}, 120:104501, 2018.

\bibitem{Meyer2017b}
Philipp Meyer and Holger Kantz.
\newblock Infinite invariant densities due to intermittency in a nonlinear
  oscillator.
\newblock {\em Phys. Rev. E}, 96:022217, Aug 2017.

\bibitem{Hurst51}
H.~E. Hurst.
\newblock Long-term storage capacity of reservoirs.
\newblock {\em Transactions of American Society of Civil Engineers}, 116:770,
  1951.

\bibitem{Peng1994}
C.~K. Peng, S.~V. Buldyrev, S.~Havlin, M.~Simons, H.~E. Stanley, and A.~L.
  Goldberger.
\newblock Mosaic organization of dna nucleotides.
\newblock {\em Phys. Rev. E}, 49:1685, 1994.

\bibitem{Hoell2015}
M.~Hoell and H.~Kantz.
\newblock The relationship between the detrendend fluctuation analysis and the
  autocorrelation function of a signal.
\newblock {\em Eur. Phys. J. B}, 88:327, 2015.

\bibitem{Kantelhardt2002}
Jan~W. Kantelhardt, Stephan~A. Zschiegner, Eva Koscielny-Bunde, Shlomo Havlin,
  Armin Bunde, and H.~Eugene Stanley.
\newblock Multifractal detrended fluctuation analysis of nonstationary time
  series.
\newblock {\em Physica A}, 316:87--114, 2002.

\bibitem{Korabel2009}
Nickolay Korabel and Eli Barkai.
\newblock Pesin-type identity for intermittent dynamics with a zero lyaponov
  exponent.
\newblock {\em Phys. Rev. Lett.}, 102:050601, 2009.

\bibitem{Dynkin}
Evgeniĭ~Borisovich Dynkin.
\newblock {\em Selected Translations in Mathematical Statistics and
  Probability}, 1:171, 1961.

\bibitem{Thaler2005}
Maximilian Thaler.
\newblock Asymptotic distributions and large deviations for iterated maps with
  an indifferent fixed point.
\newblock {\em Stoch. Dyn.}, 5:425, 2005.

\bibitem{AkimotoBarkai}
Takuma Akimoto and Eli Barkai.
\newblock Aging generates regular motion in weakly chaotic systems.
\newblock {\em Phys. Rev. E}, 87:032915, 2013.

\bibitem{Beck1987}
C.~Beck and G.~Roepstorff.
\newblock Effects of phase space discretization on the long-time behavior of
  dynamical systems.
\newblock {\em Physica D: Nonlinear Phenomena}, 25(1):173 -- 180, 1987.

\bibitem{Grebogi1988}
Celso Grebogi, Edward Ott, and James~A. Yorke.
\newblock Roundoff-induced periodicity and the correlation dimension of chaotic
  attractors.
\newblock {\em Phys. Rev. A}, 38:3688--3692, Oct 1988.

\bibitem{BasslerPhysicaA06}
Kevin~E. Bassler, Gemunu~H. Gunaratne, and Joseph~L. McCauley.
\newblock Markov processes, hurst exponents, and nonlinear diffusion equations:
  With application to finance.
\newblock {\em Physica A: Statistical Mechanics and its Applications},
  369(2):343 -- 353, 2006.

\bibitem{McCauleyPhysicaA07}
Joseph~L. McCauley, Gemunu~H. Gunaratne, and Kevin~E. Bassler.
\newblock Hurst exponents, markov processes, and fractional brownian motion.
\newblock {\em Physica A: Statistical Mechanics and its Applications}, 379(1):1
  -- 9, 2007.

\bibitem{McCauleyPhysicaA08}
Joseph~L. McCauley, Kevin~E. Bassler, and Gemunu~H. Gunaratne.
\newblock Martingales, detrending data, and the efficient market hypothesis.
\newblock {\em Physica A: Statistical Mechanics and its Applications},
  387(1):202 -- 216, 2008.

\bibitem{Mandelbrot1963}
Benoit Mandelbrot.
\newblock The variation of certain speculative prices.
\newblock {\em The Journal of Business}, 36(4):394--419, 1963.

\bibitem{MantegnaStanley1995}
Rosario~N. Mantegna and H.~Eugene Stanley.
\newblock Scaling behaviour in the dynamics of an economic index.
\newblock {\em Nature}, 376:46, 1995.

\end{thebibliography}
\end{document}